\documentstyle[aas2pp4]{article}

\begin{document}

\title{EVOLUTION OF CLUSTER ELLIPTICALS AT $0.2 < z < 1.2$
FROM  HUBBLE SPACE TELESCOPE IMAGING\footnotemark[1]$^,$\footnotemark[2]}

\author{David Schade,  L. Felipe Barrientos, \& and Omar L\'{o}pez-Cruz\footnotemark[3]$^,$\footnotemark[4]} 
\affil{Department of Astronomy, University of Toronto, 60 St. George
St., Toronto, Canada M5S 3H8}

\footnotetext[1]{Based on observations with the NASA/ESA {\it Hubble Space 
Telescope}
obtained at the Space Telescope Science Institute, which is operated by the
Association of Universities for Research in Astronomy Inc., under NASA
contract NAS 5-26555. }
\footnotetext[2]{Based on data obtained through the facilities
of the Canadian Astronomy Data Centre. }
\footnotetext[3]{{\it also} Instituto Nacional de Astrof\'{\i}sica 
Optica y Electr\'onica (INAOE), Tonantzintla, M\'exico}
\footnotetext[4]{Visiting Astronomer, Kitt Peak
National Observatory, National Optical Astronomy Observatories, which
is operated by the Association of Universities for Research in
Astronomy, Inc., under cooperative agreement with the National Science
Foundation.}

\begin{abstract}

Two-dimensional surface photometry derived from {\em Hubble Space 
Telescope} imaging is presented for a sample
of 225 early-type galaxies (assumed to be cluster members) in the fields of 9 clusters at
redshifts $0.17 < z < 1.21$. 
The 94 luminous ellipticals 
($M_{AB}(B)<-20$; selected by morphology alone with no 
reference to color) form  tight sequences in
the size-luminosity plane. 
The position of these sequences shifts, on average, with redshift
so that an object of a given size at $z=0.55$ is brighter by
$\Delta M(B)=-0.57 \pm 0.13$ mag than its counterpart 
(measured with the same techniques) in nearby clusters.
At $z=0.9$ the shift is $\Delta M(B)=-0.96 \pm 0.22$
mag. If the relation between size
and luminosity is universal so that the local cluster
galaxies represent the evolutionary endpoints of those at
high redshift, and if the size-luminosity relation
is not modified by dynamical processes then
this population of galaxies has undergone significant luminosity evolution
since $z=1$ consistent with expectations based on models
of passively evolving, old stellar populations.

\end{abstract}


\section{INTRODUCTION}

 Luminosity evolution is an expected consequence of the
passive aging of a stellar population 
 such as that believed 
to make up the bulk of the stars in
elliptical galaxies (e.g., Tinsley 1972).  Signs of 
color evolution (which accompanies
luminosity evolution)  
have been reported by Dressler and Gunn (1990),  
Aragon-Salamanca et al. (1993), Rakos and Schombert (1995),
and Oke, Gunn, \& Hoessel (1996).
These observations are all broadly consistent with theoretical
models of
old, passively-evolving, elliptical galaxies (e.g., Bruzual \& Charlot 1993).

 Observations with {\em Hubble Space Telescope} by Pahre, Djorgovski, 
\& de Carvalho (1996) of a cluster at $z=0.41$
indicate evolution relative to the local Kormendy (1977) relation
of $-0.36 \pm 0.14$ 
magnitudes in the restframe $K$-band and Barrientos, 
Schade, \& L\'{o}pez-Cruz (1996) 
find $-0.64\pm 0.3$ mag of evolution in the size-luminosity 
relation in that same cluster relative to Coma.
 Spectroscopic observations by  Bender, Ziegler, \& Bruzual (1996)
and van Dokkum \& Franx (1996) result in fundamental plane 
and Faber-Jackson relations at $z\sim 0.4$ that are both consistent with the
photometric evidence for evolution.

 Schade et al. (1996a) use two-dimensional modeling techniques 
to analyze a sample of 166 early-type galaxies and 
show that cluster and field ellipticals
evolve with redshift in the $M_B - \log R_e$ plane (a projection of the fundamental
plane). If the $M_B - \log R_e$ relation is
not significantly modified by dynamical processes then this is 
the signature of  luminosity evolution of individual galaxies
amounting to $\sim -0.5$ mag at $z=0.5$. 
{\em HST} data yields more precise surface photometry because
of its superior resolution. Archival imaging is 
available for a number of clusters although  
the {\em HST} field typically
covers less than 1 Mpc$^2$ so the number of luminous galaxies
in a single pointing is small and, furthermore, few redshifts are available. 
Nevertheless, this approach is
complementary to that of Schade et al. (1996a) as a means of detecting
the evolution of cluster galaxies. 

This {\it Letter }  describes the analysis of 
two-dimensional surface photometry of a sample of 94  luminous 
{\em early-type} galaxies in the fields of 9 clusters at $0.17 < z < 1.21$ .
Data and procedure  are  described in \S 2. 
The relation between size and luminosity
or surface brightness is presented in \S 3 and the
results are discussed in \S 4.
 It is assumed throughout this paper that $H_\circ=50$ km sec$^{-1}$ Mpc$^{-1}$ 
and $q_{\circ}=0.5$.

\section{DATA AND PROCEDURE}

 {\em HST} imaging of 9 
galaxy clusters with $0.17 < z <  1.21$ was obtained from the {\em HST} archive
using the facilities of the Canadian Astronomy Data Centre.
Table 1 gives details of the data. 
 Object catalogs were constructed by eye for all of
the clusters and an additional matched-template finding algorithm was applied to those
clusters at $z > 0.5$. Elliptical galaxy templates with a variety of sizes,
axial ratios, and orientations (90 templates in total) were convolved
with the images of the $z > 0.5$ clusters and a detection significance
was computed at each pixel of the image. All
detections with signal-to-noise ratio greater than 30 were selected
and subjected to the fitting procedure.
This finding procedure allows the selection limits to be determined
so that the limiting surface brightness for detection of an elliptical galaxy
of a given size can be computed.
Selection lines are shown
on Figure 2. The galaxy locus in
the $M_B -\log R_e$ plane lies well away from the selection lines
except at the highest redshift. Therefore
the photometric catalogs are complete in these clusters down to luminosities
(and surface brightnesses) well below the sequence of luminous galaxies that is
used to derive the shifts in the $M_B-\log R_e$ relation.

Two-dimensional models (exponential disks and
$R^{1/4}$ laws) were fitted to the symmetric components of
the light distributions of 
galaxies (see Schade et al. 1995) in order to
minimize the effects of nearby companions. Galaxy models were 
convolved with the empirically-determined
point-spread function  constructed 
using the IRAF task DAOPHOT (Stetson 1987). 
There are typically few bright, unsaturated stars
on a WFPC2 frame from which to construct an effective
point-spread function (PSF) so a standard WFPC2 PSF
constructed from  6 bright stars was adopted for most clusters. 
The frame-to-frame variations in the PSF (and spatial variations
within a frame) contribute about 5\% r.m.s. noise to the
size measurments when a standard PSF is used.
In the 
few cases where enough stars were available, individual-frame
PSFs were constructed. 

The idealized galaxy models were convolved with the
PSFs  and
a $\chi^2$ minimization routine was used to find the best-fit
model parameters (size, surface brightness,
and fractional bulge luminosity, $B/T$). The error bars shown
Figures 1 and 2 are based on output from the fitting routine and are
confirmed by simulations to be reasonable in the presence of
sky-subtraction and other random errors. 
Details of the
fitting procedure are given in Schade et al. (1996b).
Those objects that were 
fit well by a single component $R^{1/4}$ law (as judged by this
visual inspection of the residuals) were defined as
elliptical galaxies for the purposes of the present work. 
No color information was used in the selection
process. 
A total of more than 1500 objects were
fitted and 225 were classified as ellipticals. 

 The photometric zeropoints from the WFPC2 image headers were used
to convert counts to flux densities (Whitmore 1995) and then 
to $AB$ magnitudes.
The $AB$ magnitudes were converted to restframe $M_{AB}(B)$
[$B_{AB}=B-0.17$] luminosities
 using interpolation of a {\em present-day} elliptical galaxy 
spectral energy distribution (Coleman, Wu, and Weedman 1980)
as described by Lilly et al. (1995). The K-correction problem
is discussed in \S 3. 

\section{RESULTS}

 An important feature of this study is that the imaging of nearby clusters
(obtained with the Kitt Peak 0.9 m telescope [L\'{o}pez-Cruz  1996])
which provides the local anchoring point for the $M_B-\log R_e$ relation 
has similar physical resolution 
and signal-to-noise ratio to the $HST$ data (see Barrientos et al. 1996)
and were measured with identical techniques. Thus these $B$-band measures form a
consistent comparison set for the high-redshift galaxies.

Figure 1 shows  the relation between $\log R_{e}$ (half-light radius
in kpc) and $M_{AB}(B)$ for elliptical galaxies in 4 nearby clusters.
 Extinction corrections in the $B$-band of
0.18 mag for A2256 ($z=0.0601$), 0.07 mag for A2029 ($z=0.077$), 0.05
mag for Coma ($z=0.023$), and 0.04 for A957 ($z=0.045$) were obtained
from the Nasa Extragalactic Database which provides Burstein and Heiles
(1982) values with relative errors in $E(B-V)$ estimated at 0.01 mag.
 With these corrections the 
$M_B -\log R_e$ relations for these 4 clusters are consistent with one
another (the offset in luminosity between A2256 and A2029 which have 38 and 60
elliptical galaxies brighter than $M_{AB}(B)=-20$ is $0.00 \pm 0.10$ mag)
and they were combined to form the adopted local relation: 
$M_{AB}(B)=-3.33 \log R_e - 18.65\pm 0.06$.
The slope was constrained to agree
with Schade et al. (1996a) and the present relation supercedes that work. 
It fits all of the present data well
and the computed offsets between this local relation and those at
high redshift is insensitive to this choice.

Figure 2 shows  the relation between $\log R_{e}$ (half-light radius
in kpc) and $M_{AB}(B)$ for elliptical galaxies in the
fields of 9 clusters at $ 0.17 < z < 1.21$. 
Luminosities and sizes are calculated
assuming that all galaxies are cluster members. Estimates
of contamination by field ellipticals from the counts of Driver et al. (1996)
show that contamination of the bright sequence of cluster galaxies ($M_B(AB) < -20$)
 is less than 1 or 2 galaxies for clusters at $z < 0.55$ although the
corrections for higher redshift clusters become significant (e.g., 9 contaminating
galaxies in 3C324). However, plots of observed parameters $R_e$ in arcseconds
versus apparent magnitude show that the ellipticals in the 3 fields
occupy discrete bands. This is a strong indication that the ellipticals
are indeed predominantly cluster sequences. [There is a second cluster
sequence of 6 galaxies with an estimated redshift of 0.45 in the field
of 3C324.] Although some field contamination
is likely to be present, it must be smaller than suggested
by the counts of Driver et al. (1996) and does not dominate the
evolutionary results derived here. 

On average, the $M_B -\log R_e$ sequences in Figure 2 shift toward higher luminosity
with increasing redshift. The size of the shifts and corresponding errors were computed using
the techniques given in Feigelson \& Babu (1992) and also computed according
to the same procedure used in Schade et al. (1996a). The 
estimated shifts and errors were virtually identical these two techniques and
the results presented here were computed adopting a slope of 
$\Delta M/ \Delta log R_E=-3.33$ to agree with Schade et al. (1996a). 
Only those galaxies
with $M_{AB}(B)< -20$ were included in the computation
of $\Delta M$.
The two clusters common to the present study and Schade et al.
(1996a) have independently
measured values of $\Delta M(B)$ (after revisions discussed in \S 4) 
that differ by $0.12 \pm 0.21$
mag [ABELL2390] and $0.03 \pm 0.24$ mag [CL001558+16].

The adoption of a
present-day elliptical spectral energy distribution 
to compute the restframe $B$-band (4400 \AA) luminosity 
results in an underestimate (by $\sim 0.15$ mag at $z=1$ and
less at lower redshift [Charlot, Worthey, \& Bressan 1996])
of the $B$-band luminosity
of a younger (bluer) stellar population if 
the observations are at a restframe wavelength longer than
4400 \AA. 
 At z=0.751 and z=0.895 the observed wavelengths are 4576 and
$4230 \AA$ respectively in the rest frame
 so that the uncertainties in the K-corrections are 
small ($< 0.05$ mag). The situation is more difficult
at z=1.206 where the observed wavelength (F702W) corresponds to
the ultraviolet ($3160 \AA$) so that this point is much 
less secure.  
Dickinson (1995) suggests that these galaxies, although still very red,
are $\sim 0.6$ mag bluer in $R-K$ than a present-day
elliptical.  The models of Bruzual \& Charlot (1993) predict a change of only
$\sim 0.4$ mag from t=5 Gyr to t=15 Gyr (corresponding roughly
to $z=1$ and the present day). Evolutionary corrections would
would decrease the amount of brightening measured at $z=1.206$. 

 The effect of cosmology on this result is indicated by the arrows in the lower
left of each cluster panel. These show the change in size and luminosity that
result from changing $q_\circ=0.5$ to $q_\circ=0.1$. The net effect
on the computed magnitude shifts is reduced by the fact that much of the
change is parallel to the $M_B-\log R_e$ relation. For $q_\circ=0.1$ 
stronger evolution would be computed by $-0.06$ mag at $z=0.4$,
$-0.09$ mag at $z=0.55$, and $-0.17$ mag at $z=1.2$.

\section{DISCUSSION}

 The relationship between blue luminosity and effective radius
for elliptical galaxies in the present sample shifts, on average,
with redshift so that at $z=0.9$ a galaxy of 
a given size is $-0.96 \pm 0.22$ mag more luminous than
its counterpart (of the same size) in our sample
of local clusters. 
If the $M_B- \log R_e$ relation for elliptical
galaxies is universal and if dynamical
evolution does not modify the $M_B- \log R_e$ relation (e.g., Capelato, de 
Carvalho, \& Carlberg 1995) then this
effect is due to luminosity evolution of individual
galaxies.

 The galaxies in Figure 1 were selected on the basis of their morphology
alone with no reference to their color. Only those galaxies that are well-fit
by an $R^{1/4}$ law were classified as ellipticals. 
This selection process was done by visual inspection and
is thus subjective although the position of the galaxies on the
$M_B-\log R_e$ diagram was not known at the time that the classifications
were done.  The selection criteria ensure that all of the
galaxies included in the present study conform
closely to an $R^{1/4}$ law, but some galaxies may have been rejected
that would be classified as elliptical galaxies using, e.g., color 
criteria. 
 The main weakness in the present study is the lack of spectroscopic
information so that cluster membership is uncertain. We believe that
field galaxy contamination is small (see \S 3) but these results
would clearly be much more secure if these galaxies were confirmed
cluster members. 

 A number of studies detect evolution in the colors of the
reddest galaxies in clusters. 
Aragon-Salamanca et al. (1993), 
 Rakos \& Schombert (1995),  and Oke, Gunn, \& Hoessel (1996)
detect color changes that are consistent with old, single-burst models 
of elliptical galaxies.
 Spectroscopic studies, although based on
small numbers of galaxies, support the conclusions of the photometric work.
Bender, Ziegler, \& Bruzual (1996) use velocity dispersions
and Mg line strengths of 16 galaxies to derive evolution 
of $\sim 0.5$ mag in a cluster at $z=0.37$ and van Dokkum \& Franx (1996)
detect a few tenths of a magnitude of evolution in the fundamental
plane at $z=0.39$ from observations of 9 galaxies. All of these
results imply that the luminosities of ellipticals should be 
measurably larger at $z >0.4$.

 Figure 3 shows the evolution with redshift of the $M_B- \log R_e$
relation for elliptical galaxies. $\Delta M$ is the change
in luminosity (or, equivalently, surface brightness) for a galaxy
of a given size. Included are results from Schade et al. (1996a)
for field and cluster galaxies and from
Barrientos et al. (1996). 
The best-fit relation is $\Delta M =
0.78 \Delta z$.
Superimposed upon Figure 3
is the predicted brightening from Buzzoni (1995) for three
values of the initial mass function (IMF) and an assumed age of
15 Gyr at the present time.
 Note that these theoretical
tracks are $\Delta M(V)$ rather than $B$-band values 
which would be $\sim 0.15$ mag larger  at $z=1$ (Charlot, Worthey, 
\& Bressan 1996). If the difference in bands is ignored then
a best-fit IMF slope is $s=2.85\pm0.2$ (where a Salpeter slope
is $s=2.35$). 

It has been shown (Dressler \& Gunn 1992) that
13\% of the galaxies with deVaucouleurs ($R^{1/4}$) profiles 
in their set of clusters
at $z\sim 0.4$ show spectroscopic evidence of recent star-formation
(i.e., have ``E+A spectra'' showing Balmer absorption). We expect 
some fraction of our own sample to have E+A spectral energy
distributions and such galaxies would contribute to the observed evolution
of the $M_B- \log R_e$ relation. If our sample of
ellipticals were dominated by such galaxies at high redshift
then the observed evolution in the $M_B- \log R_e$
would be strongly affected by recent star-formation. 

 The improved local calibration of the $M_B -\log R_e$ relation given above
results in a downward revision of the amount of evolution found 
for cluster and field galaxies in Schade et al. (1996a)
by 0.02, 0.14, and 0.13 mag for A2390, MS1621+26, and MS0016+16.
$B$-band galactic extinctions are 0.18 mag (A2256),
0.33 mag (A2390), 0.08  (MS1621+26) and 0.11 mag (MS0016+16).
All extinctions were obtained from the Nasa Extragalactic Database.
The  Gunn $r$ extinction is about half of the $B$-band
extinction (Mihalas \& Binney 1981). 
In Schade et al. both the local calibration and the high-redshift
cluster photometry were zero-pointed against PPP (Yee 1991) photometry
whereas for the present calibration the total galaxy magnitude is
obtained from the fitted model parameters (thus the galaxy light
is integrated to infinity).
The difference amounts to 10-15\%.

The evolution found in the present work is consistent with that found by
Schade et al. (1996a) to $z=0.55$ and shows that the
evolutionary trend continues to $z\sim 1$.   
Schade et al. detect no significant difference 
between field and cluster elliptical galaxy evolution although
it is important to note that their  
cluster sample is dominated by galaxies in regions
far from the cluster core. In contrast, the present study is restricted to
galaxies less than 1 Mpc from the core. The fact that we see no difference
in $\Delta M_B$ between these galaxies near the cluster core, those in
the outer regions of the cluster, and those in the field, is further
evidence for the homogeneous nature of the elliptical galaxy population,
(e.g., Bower, Lucey \& Ellis 1992). Neither the $M_B-\log R_e$ relation
nor its evolution has yet revealed a significant dependence 
on environment.

\begin{acknowledgments}

 We thank Simon Lilly, Ray Carlberg, and Pedro Colin for helpful discussions. 
 We also thank an anonymous referee for suggestions which greatly
improved this paper.
We thank Eric Feigelson for kindly providing his computer code.
The assistance of the Canadian Astronomy Data Centre 
is very much appreciated. 
 This work was supported financially by NSERC of Canada.

\end{acknowledgments}

\newpage
\centerline{Figure Captions}

\begin{figure}
\caption{The relations between $M_{AB}(B)$ and $\log R_e$ (half-light
radius in kpc) for elliptical galaxies in nearby clusters. These  four
clusters together form the adopted local relation against which
the high-redshift results are compared.}

\end{figure}

\begin{figure}
\caption{ The relation between $M_{AB}(B)$ and $\log R_e$ 
for elliptical galaxies in moderate and high-redshift clusters. The solid line in
all panels corresponds to the best-fit to the local sequence 
(upper left panel) defined in Figure 1 
and the dashed lines indicate the best-fit fixed-slope
($\Delta M_B/\Delta \log R_e=-3.33$) relation for each cluster.
The fits were restricted to those galaxies with $M_{AB}(B) < -20$
in all clusters. Arrows indicate the effect of changing $q_\circ=0.5$
to $q_\circ=0.1$. For those clusters at $z > 0.5$ we also show the selection
line (signal-to-noise ratio greater than 30 from the matched-template finding
algorithm described in the text) as a dashed line in the upper-right corner
of those panels. Elliptical galaxies to the left of that line will 
be detected.}

\end{figure}

\begin{figure}
\caption { The luminosity shift $\Delta M_B$  including revised results from
Schade et al. (1996a) and from Barrientos et al. (1996). Filled symbols
are for clusters in the present study, open circles are cluster points
from Schade et al. (1996a)
(including 2 clusters in common that have been offset in redshift
slightly for clarity) and open squares are field galaxies from
Schade et al. (1996a). The lines are from the models of Buzzoni (1995)
for IMF power-law indices of $s=3.35$ (solid line), $s=2.35$ (short dashed),
and $s=1.35$ (long dashed). The models assume a present-day age of 15 Gyr
and we show the theoretical $\Delta M(V)$ whereas the data is in $M_B$ where the evolution
would be $\sim 0.15$ mag larger by  $z=1$.  }

\end{figure}

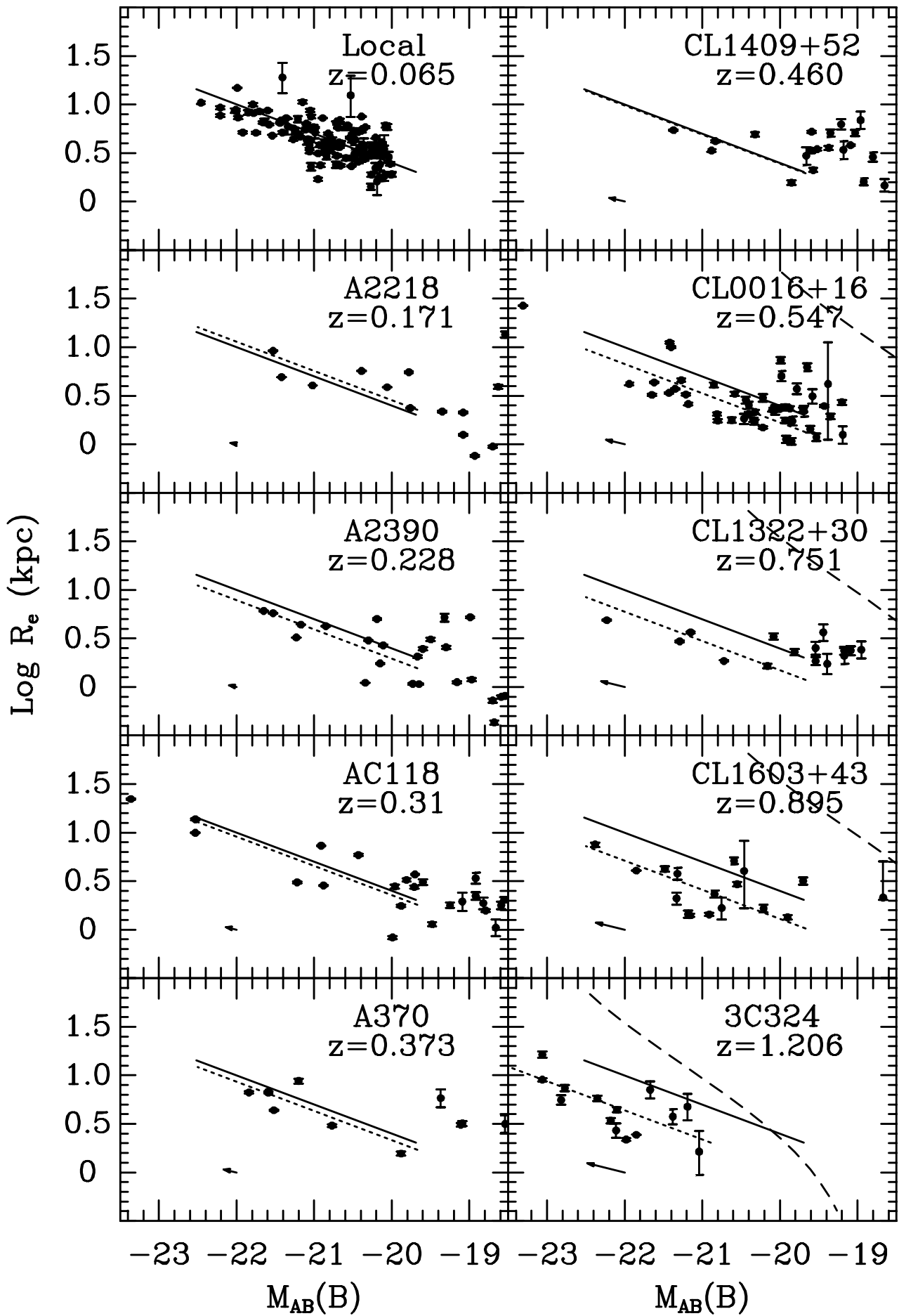
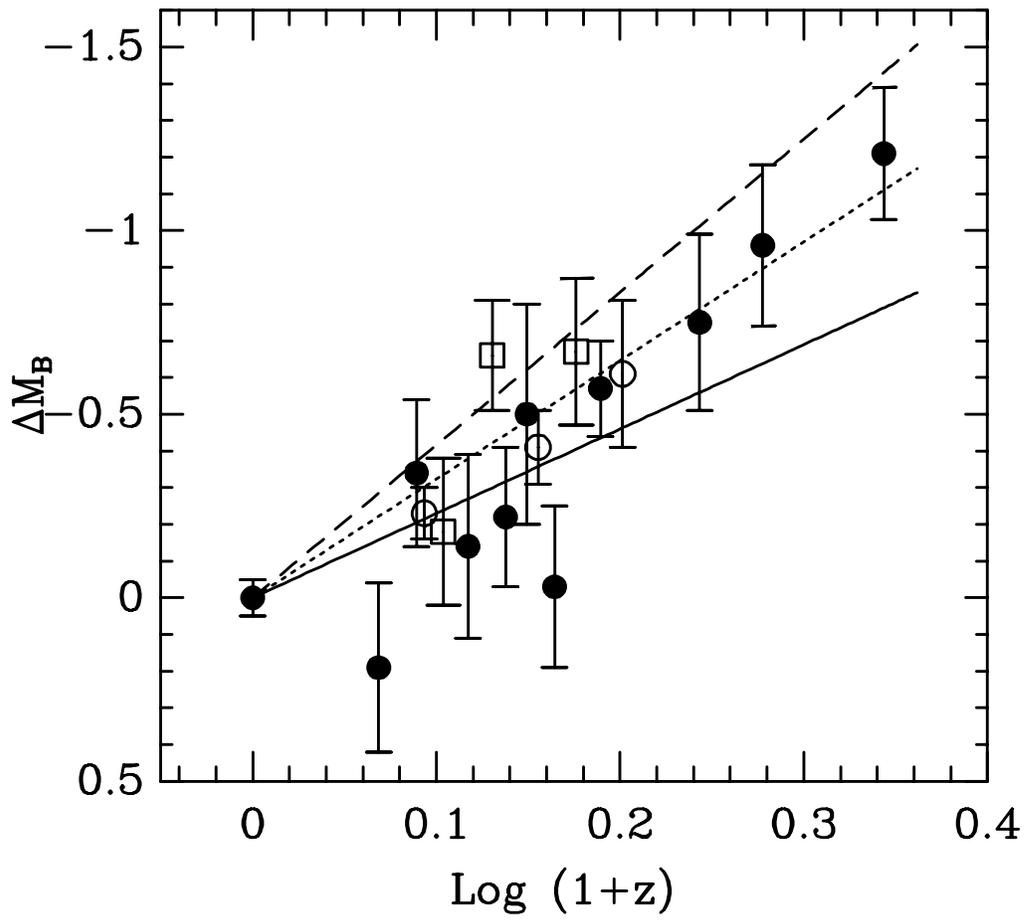
\end{document}